\documentclass[5p,twocolumn]{elsarticle}

\usepackage[dvipdfmx]{hyperref}
\usepackage{amsmath, amssymb}
\usepackage{lineno}
\modulolinenumbers[5]
\usepackage{txfonts}

\journal{Physics Letters B}

\begin{document}

\begin{frontmatter}

\title{A classical mechanical model of two interacting massless particles in de Sitter space and its quantization}

\author{Naohiro Kanda}
\ead{naohiro.kanda@gmail.com}
\author{Satoshi Okano\corref{cor1}}
\ead{satoshi.okano15@gmail.com}

\address{Research Institute of Science and Technology, College of Science and Technology, Nihon University, Chiyoda-ku, Tokyo 101-8308, Japan}

\begin{abstract}
A conformally invariant model of two interacting massless
particles in Minkowski space was proposed by Casalbuoni and Gomis
\cite{Casalbuoni2014c}.
We generalize this model to the case of de Sitter space from the perspective of geodesic distance, in such a way that the resulting, generalized action reduces to the original action in a limit that de Sitter radius goes to infinity.
We analyze the Hamiltonian formulation in accordance with Dirac's prescription for
constrained Hamiltonian systems and carry out its subsequent canonical quantization in coordinate representation following DeWitt.
As the result, we derive a fourth-order differential wave equation for bilocal fields that, in the infinite radius limit, reproduces one obtained in the original model for Minkowski space case.
\end{abstract}

\end{frontmatter}

\linenumbers

\section{Introduction}

The early universe during cosmological inflation can be approximated as de Sitter space \cite{Kolb:1990vq,Imponente2004,Brandenberger2011}.
On the other hand, the universe after inflation approaches  Minkowski space, which can be regarded as a limit that de Sitter radius $R$ goes to infinity.
During inflation and for a while after its end, the energy scale of the universe  remains much higher than the electroweak scale.
In the period, except for the Higgs particle, all the elementary particles included in the standard model  remain  massless.
In particular, after inflation, their dynamics should have conformal invariance in Minkowski space.
However, until the end of inflation, during which the arena of physics is de Sitter space, it is not clear whether conformal invariance is realized.
This observation motivates us to consider a massless system that does not have conformal invariance at the beginning but restores it in the limit $R \rightarrow \infty$.
In this letter, we construct such a model at classical and quantum mechanical levels.

A conformally invariant model of interacting massless particles in Minkowski space was given by Casalbuoni and Gomis \cite{Casalbuoni2014c}.
%%%%%
An action for two free massless particles in Minkowski space is 
\begin{align} \label{S0}
S_0 = \int d \tau L_0 , \; \quad L_0 = - \sum\limits^{2}_{i=1} \frac{ \eta_{\mu \nu} \dot{x}^{\mu}_{i} \dot{x}^{\nu}_{i} }{ 2 e_{i} } 
\: ,
\end{align}
where $ \mu, \nu = 0,1,2,3 $, $\eta_{\mu \nu} = \eta^{\mu \nu} = \mathrm{diag}(+,-,-,-)$, $ x^{\mu}_i (\tau) ,  i =1,2$ are the spacetime coordinates for the two particles, $e_i(\tau)$ are einbeins,  $\tau $ is the worldline parameter common to the two particles and $\dot{x}^{\mu}_i = dx_i^{\mu}/d\tau $.
The free action $S_0$ is invariant under both reparameterization of $\tau$ and conformal transformations in Minkowski space.
%%%%%
An interaction term $S_{\mathrm{int}}$ that has the same invariance as $S_0$ and does not include the velocities $\dot{x}_i$ is uniquely determined  \cite{Casalbuoni2014c}  as
\begin{align} \label{int}
S_{\mathrm{int}} = \int d \tau V (r) \:, \quad V(r) = - \frac{ \alpha^2}{4} \frac{ \sqrt{ e_1 e_2 } }{ r^2 }  \: , 
\end{align}
where $\alpha$ is the coupling constant, $r^{\mu} := x_1^{\mu} - x_2^{\mu} $ are the relative coordinates and $r^2 = \eta_{\mu \nu} r^{\mu} r^{\nu}$.
%%%%%
%%%%%
Thus, the conformally invariant action for two interacting massless particles is given as
\begin{align} \label{S}
 S:= S_0 + S_{\mathrm{int}}   %&:= S_0 + S_{\mathrm{int}} \nonumber \\
 = \int d\tau \left( -\sum^{2}_{i=1} \frac{1}{2} \frac{\eta_{\mu \nu} \dot{x}^{\mu}_{i} \dot{x}^{\nu}_{i} }{e_{i}}  -\frac{ \alpha^2}{4} \frac{ \sqrt{ e_1 e_2 } }{ r^2 } \right)  \: .
\end{align}
%%%%%
In the Hamiltonian formulation based on this action $S$, first-class constraints associated with the reparameterization invariance occur.
After canonical quantization, they are replaced with the physical state conditions and read as a fourth-order differential equation:
\begin{align} \label{waveeq-cg}
\left( \square_1 \square_2   - \frac{\alpha^4 }{16 r^4 } \right)
\varphi (x_1,x_2) = 0  \: ,
\end{align}
where $\square_{i} =   \partial^2 / \partial x_{i \mu} \partial x^{\mu}_i$ and $\varphi(x_1,x_2)$ is a wave function of the system, which we call a bilocal field in the sense that the wave function depends on two spacetime points.
%In this letter, we call a wavefunction that depends on only two spacetime points, like $\varphi(x_1,x_2)$, a bilocal field.
Our purpose is to generalize the action $S$ and its consequent wave equation to the case of de Sitter space.

In generalizing the action \eqref{S}, we rely on the fact that the squared relative coordinates $r^2 $ are considered as a special case of a squared geodesic distance.
A geodesic distance $\sigma(x_1,x_2) $ between two points $x_1^{\mu} (\tau) $ and $x_2^{\mu} (\tau)$ in a general spacetime endowed with a metric $g_{\mu \nu } (x) $ is given by 
%%%%%
\begin{align} \label{defsigma}
\sigma(x_1 , x_2) = \int_{s_1}^{s_2} d s \sqrt{ \pm g_{\mu \nu } (x) {x}'^{\mu} {x}'^{\nu} } 
\: ,
\end{align}
with $x_1^{\mu} (\tau) = x^{\mu}(s_1) ,\; x_2^{\mu} (\tau) = x^{\mu} (s_2) $, where $x^{\mu}(s)$ is the geodesic curve joining the two points, $s $ is a parameter of the geodesic connecting the two points, $x'^{\mu} := dx^{\mu} /ds$, and the positive (negative) sign is chosen if the geodesic is timelike (spacelike).
In particular, $ \sigma^2(x_1, x_2) =  \pm r^2 $ holds for Minkowski space.
%%%%%
Hence, the system described by the action \eqref{S} can be interpreted as a system of two particles  interacting with each other along the geodesic connecting them.
%%%%%
(For example, such a 2-particle system in a curved spacetime has been considered for the shockwave background in Ref. \cite{Kanda2015}.)

From this point of view, we generalize the action \eqref{S} to a general curved background as
\begin{align} \label{S'}
 S_{g} %&:= S_0 + S_{\mathrm{int}} \nonumber \\
 &= \int d\tau \left( -\sum^{2}_{i=1} \frac{1}{2} \frac{ g_{\mu \nu} (x_i) \dot{x}^{\mu}_{i} \dot{x}^{\nu}_{i} }{e_{i}}  \mp \frac{ \alpha^2}{4} \frac{ \sqrt{ e_1 e_2 } }{ \sigma^2(x_1,x_2) } \right)  \: , 
\end{align}
where the negative (positive) sign is chosen if the two particles are timelike (spacelike) sepalated.
%%%%%
For a spacetime such that  $\sigma^2 \rightarrow \pm r^2 $ as $g_{\mu \nu} \rightarrow \eta_{\mu \nu}$, this action smoothly reduces to the conformally invariant action \eqref{S} in Minkowski space.
However, it is noted that the generalized action \eqref{S'} itself does not necessarily have conformal symmetry before taking such a limit.

This letter is organized as follows: In the next section, we write the action $S_g$ for the case of 4-dimensional de Sitter space by using embedding coordinates, commonly used to describe (anti-)de Sitter space. 
(See for example \cite{Kuzenko1995, Dorn2006}.)
This description enables us to express the geodesic distance in a simple form without taking a particular coordinate system of de Sitter space.
%%%%%
In Sec.~3, we consider the Hamiltonian formulation in accordance with Dirac's procedure \cite{Dirac:1964:LQM} for constrained systems.
%%%%%
In Sec.~4, new canonical variables constituting canonical pairs in the sense of Dirac brackets are introduced.
%%%%%
In Sec.~5, we apply the canonical quantization procedure in coordinate representation for curved configuration spaces following DeWitt \cite{Dewitt1952}.
Sec.~6 is devoted to conclude this letter.

\section{Lagrangian formulation}

In this section, we specify the form of the action \eqref{S'} for the case of a 4-dimensional de Sitter space in terms of embedding coordinates and then consider its limit $R \rightarrow \infty $.

A $4$-dimensional de Sitter space $\mathrm{dS}_4$ of radius $R$ can be embedded into a $(1+4)$-dimensional flat space $\mathbb{R}^{1,4}$ as a hyperboloid
\begin{align} \label{dS}
\mathrm{dS}_4 = \{ X^A \in \mathbb{R}^{1,4} \; | \; \eta_{AB} X^A X^B = - R^2  \:  \} \: ,
\end{align}
where $X^A \: (A = 0,1,2,3 ,5)$ are embedding coordinates and $ \eta_{AB} = \mathrm{diag} ( +1, -1, -1, -1, -1)$.
This hypersurface has a parameterization 
\begin{align} \label{local}
X^{\mu} = x^{\mu} \: , \quad X^{5} = \pm \sqrt{ \eta_{\mu \nu} x^{\mu} x^{\nu} +R^2 }
\end{align}
with $\mu , \nu  = 0,1,2,3$. 
Then the induced metric on the de Sitter space for this parameterization has the form 
\begin{align} \label{induced}
g_{\mu \nu} = \eta_{\mu \nu} - \frac{ x_{\mu } x_{\nu} }{ \eta_{\alpha \beta} x^{\alpha} x^{\beta} +R^2 } \: , 
\end{align}
which approaches the 4-dimensional Minkowski metric $\eta_{\mu \nu}$ in the limit that $R \rightarrow \infty$.

In terms of embedding coordinates, a Lagrangian for two free massless particles in de Sitter space is written as
\begin{align} \label{L0}
 L_{0(\mathrm{dS})} = -\sum^{2}_{i=1} \frac{1}{2}\frac{\dot{X}^{2}_{i}}{e_{i}}
-\sum^{2}_{i=1} \lambda_{i}(X^{2}_{i} +R^{2}), 
\rule[0mm]{0pt}{5mm}
\end{align}
where $ X_i^{A} = X^{A}_{i }(\tau) \: ( i=1,2 )$ are the positions of two particles and $\lambda_{i} \: (i = 1,2) $ are the multipliers.

In this letter, we restrict ourselves to the case that the particles are spacelike separated, in which the geodesic distance $\sigma$ is written in terms of the embedding coordinates as (cf. Ref.~\cite{Cunningham2017a})
\begin{align} \label{sigma}
 \sigma(X_{1},X_{2}) &= R \cos^{-1} \left( - \frac{X_{1} \cdot X_{2}
}{R^{2}} \right) 
.
\end{align}
Thus the interaction term is given as
\begin{align} \label{V}
V_{\mathrm{dS}} =  \frac{ \alpha^2}{4} \frac{ \sqrt{ e_1 e_2 } }{  \sigma^2 (X_{1},X_{2}) } \: .
\end{align}
%%%%%
From Eqs.~\eqref{L0} and \eqref{V}, the generalized action \eqref{S'} for the case of de Sitter space becomes
\begin{align} \label{SdS}
S_{\mathrm{dS}} := \int d \tau \,  L_{\mathrm{dS}} \: , 
\end{align}
with the Lagrangian is expressed as
\begin{align} \label{CL2} 
 L_{\text{dS}} &:= L_{0(\mathrm{dS})}  +  V_{\mathrm{dS}}  \nonumber \\
 &=  -\sum^{2}_{i=1} \frac{1}{2}\frac{\dot{X}^{2}_{i}}{e_{i}}
+ \frac{\alpha^{2}}{4}\frac{\sqrt{e_{1}e_{2}}}{\sigma^{2}}
 -\sum^{2}_{i=1} \lambda_{i}(X^{2}_{i} +R^{2})   \: .
\end{align}

The  action $S_{\text{dS}}$ is invariant under the reparameterization $\tau \rightarrow \tau^{\prime}(\tau) $ with the transformation rules of $X_i^A, e_i$ and $\lambda_i $:
\begin{subequations}
\begin{align}
 X^{A}_{i}(\tau) & \: \rightarrow \:  X^{\prime A}_{i}(\tau^{\prime}) = X^{A}_{i}(\tau) \:  ,\\
 e_{i}(\tau) & \: \rightarrow \: e^{\prime}_{i}(\tau^{\prime}) = \frac{d\tau}{d\tau'} e_{i}(\tau) \: ,\\
 \lambda_{i}(\tau) & \: \rightarrow \: \lambda^{\prime}_{i}(\tau^{\prime}) = \frac{d\tau}{d\tau^{\prime}} \lambda_{i}(\tau) \: .
\end{align}
\end{subequations}
The action $S_{\mathrm{dS}}$ also leaves invariant under the isometry of de Sitter space
\begin{align}
X_i^A \: \rightarrow \: X'{}_i^A = L^A{}_B X_i^B  \:  ,
\end{align}
where $e_i$ and $\lambda_i$ are left unchanged and $ (L^A{}_B) \in SO(4,1)$ is a constant matrix.
Although the action $S$, given in Eq.~\eqref{S}, is invariant under conformal transformations in Minkowski space, its generalized action $S_{\mathrm{dS}}$ for de Sitter space cannot be invariant under an infinitesimal conformal isometry \cite{Chernikov1968}
\begin{align} \label{infsml}
\delta X_i^A &= (b_B X_i^B )  X_i^A + R^2 b^A \: ,  
\end{align}
with no sum over $i$, where $b^A$ are infinitesimal constant parameters.
\footnote{
With the transformation law $\delta e_i =  2  b_A X_i^A e_i $ concerning the infinitesimal conformal isometry \eqref{infsml}, we see that the following action is conformally invariant:
\begin{align*}
\tilde{S} = &  \int d\tau  \left\{  L_{0(dS)} 
   -\frac{\alpha^{2}}{4}\frac{\sqrt{e_{1}e_{2}}}{ ( X_1 - X_2 )^2 }  \right\}
   \: .
%& \qquad  \quad 
\end{align*}
However, $(X_1-X_2)^2$ does not represent the squared geodesic distance between the two points $X_1^A$ and $X_2^A$ in de Sitter space.
}

We show that the action $S_{\mathrm{dS}} $ reproduces the action $S$ for Minkowski space in a limit as $R\rightarrow \infty $.
By using the parameterization \eqref{local} and the induced metric \eqref{induced}, the free Lagrangian \eqref{L0} is written as 
\begin{align} \label{L0dS}
L_{0 (\mathrm{dS})} = - \sum_{i=1}^2  \left( \eta_{\mu \nu} - \frac{ x_{i\mu } x_{i\nu} }{ \eta_{\alpha \beta} x_i^{\alpha} x_i^{\beta} +R^2 } \right) \frac{  \dot{x}_i^{\mu} \dot{x}^{\nu}_i }{2e_i }  \: ,
\end{align}
while the geodesic distance \eqref{sigma} can be expanded around the point $ \sqrt{-r^2} / \sqrt{2}R  = 0 $ as 
\begin{align} \label{sigmaex}
\sigma \big( X_1 (x_1) ,X_2 (x_2) \big) =  \sqrt{ - r^2} + \mathcal{O} \left( R^{-1} \right)  \:\: 
\end{align}
with  $r^{\mu} = x_1^{\mu} - x_2^{\mu} $ and $r^2 = \eta_{\mu \nu } r^{\mu} r^{\nu}  < 0$.
%%%%%
From Eqs.~\eqref{L0dS} and \eqref{sigmaex} we see that $ L_{0(\mathrm{dS})} \rightarrow \: L_0  $ and $  V_{\mathrm{dS}} \: \rightarrow \:  V(r)   $ as $R \rightarrow \infty $.
%%%%%
Therefore the action $S_{\mathrm{dS}} $ reproduces the action $S$ for Minkowski space case in the limit of the infinite de Sitter radius.

\section{Hamiltonian formulation}

In this section, we consider the constrained Hamilton system governed by the Lagrangian \eqref{CL2} in accordance with Dirac's procedure.

We treat $(X_1^A, X_2^A, e_1, e_2, \lambda_1, \lambda_2)$ as the canonical coordinates and define the canonical momenta $(P_1{}_A, P_2{}_A, P_{e_1} , P_{e_2}, P_{\lambda_1}, P_{\lambda_2})$ conjugate to them as
%%%%%
\begin{subequations} \label{P}
\begin{align}
P_1{}_A &:= \frac{\partial L_{\mathrm{dS}} }{\partial \dot{X}_1^A} = - \frac{\dot{X}_1{}_A}{e_1} \: , & 
P_2{}_A &:= \frac{\partial L_{\mathrm{dS}}}{\partial \dot{X}_2^A} = - \frac{\dot{X}_2{}_A}{e_2} \: , &  \\
P_{e_1} &:= \frac{\partial L_{\mathrm{dS}}}{\partial \dot{e}_1 } = 0 \: , \label{P3} & 
P_{e_2} &:= \frac{\partial L_{\mathrm{dS}}}{\partial \dot{e}_2 } = 0 \: , &  \\
P_{\lambda_1} &:= \frac{\partial L_{\mathrm{dS}}}{\partial \dot{\lambda}_1 } = 0 \: , & 
P_{\lambda_2} &:= \frac{\partial L_{\mathrm{dS}}}{\partial \dot{\lambda}_2 } = 0  \: .& 	
\label{P4} 
\end{align}
\end{subequations}
%%%%%
The nonvanishing Poisson brackets between all the canonical variables are
%%%%%
\begin{subequations} \label{PB}
\begin{align} 
\{ X_1^A \; , \: P_{1B} \} &= \delta^A_B \: , & \{ X_2^A \; , \: P_{2B} \} &= \delta^A_B \: ,  \\
\{ e_1 \; , \: P_{e_1} \} &= 1 \: , & \{ e_2 \; , \: P_{e_2} \} &= 1 \: , \\
\{ \lambda_1 \; , \: P_{\lambda_1} \} &= 1 \: , & \{ \lambda_2 \; , \: P_{\lambda_2} \} &= 1 \: .
\end{align}
\end{subequations}
%%%%%
From Eq.~(\ref{P}), the canonical Hamiltonian $H_{\mathrm{C}}$ is obtained by the Legendre transform of $L_{\mathrm{dS}}$ as 
%%%%%
\begin{align} \label{HC}
H_{\mathrm{C}}  := & \: P_{1A} \dot{X}_1^A + P_{2A} \dot{X}_2^A + P_{e_1} \dot{e}_1 + P_{e_2} \dot{e}_2 
\notag \\
	& 
	+ P_{\lambda_1} \dot{\lambda}_1+ P_{\lambda_2} \dot{\lambda}_2 - L_{\mathrm{dS}} 
\notag \\
=& \; - \frac{e_1}{2} P_1^2 - \frac{e_2}{2} P_2^2 
	- \frac{\alpha^2}{4} \frac{\sqrt{e_1e_2}}{ \, \sigma^2 \,}  
\notag \\
	& \: 
	+ \lambda_1 \Big( X_1^2 + R^2 \Big) + \lambda_2 \Big( X_2^2 + R^2 \Big) \: .
\end{align}

Eqs.~(\ref{P3})-(\ref{P4}) are treated as primary constraints
%%%%%
\begin{align} \label{primary}
\phi_{e_1} := \: & P_{e_1}  \approx 0 \: , & \phi_{e_2} := \: & P_{e_2}  \approx 0 \: , &  \\
\phi_{\lambda_1} := \: & P_{\lambda_1}  \approx 0 \: , & \phi_{\lambda_2} := \: & P_{\lambda_2} \approx 0 \: , &  
\end{align}
%%%%%
where the symbol ``$\approx$'' denotes weak equality and $\phi$'s are referred to as constraint quantities for primary constraints.
%%%%%
Introducing Lagrange multipliers $u_{e_1} , u_{e_2} , u_{\lambda_1} , u_{\lambda_2}$ for the primary constraints, we define the total Hamiltonian as
%%%%%
\begin{align}
H_{\mathrm{T}} := \: & H_{\mathrm{C}} + u_{e_1} \phi_{e_1} + u_{e_2} \phi_{e_2} + u_{\lambda_1} \phi_{\lambda_1} + u_{\lambda_2} \phi_{\lambda_2}  \: .
\end{align}
The $\tau$ evolution of a function $f$ of the canonical variables is governed by 
\begin{align}
\dot{f}  = \big\{ f \, , \, H_{\mathrm{T}} \big\} \: .
\end{align}

Using this equation and Eq.~(\ref{PB}), the $\tau$ evolution of the constraint quantities for primary constraints are evaluated as 
\begin{subequations}\label{PB_phiHT}
\begin{align} 
\dot{ \phi }_{e_1} &= \frac{P_1^2}{2} + \frac{\alpha^2}{8} \sqrt{ \frac{e_2}{e_1}  } \frac{1}{\sigma^2} \: , \\
\dot{ \phi }_{e_2} &= \frac{P_2^2}{2} + \frac{\alpha^2}{8} \sqrt{ \frac{e_1}{e_2}  } \frac{1}{\sigma^2} \: , \\
\dot{ \phi }_{\lambda_1} &= - \Big( X_1^2 + R^2 \Big) \: , 
\\
\dot{ \phi }_{\lambda_2} &= - \Big( X_2^2 + R^2 \Big) \: . 
\end{align}
\end{subequations}
The consistency conditions $(\dot{\phi}_{e_1} , \dot{\phi}_{e_2}, \dot{\phi}_{\lambda_1}, \dot{\phi}_{\lambda_2} ) \approx 0 $ require secondary constraints:
\begin{subequations}\label{secondary}
\begin{align} 
\chi_{e_1 } &:=\:    P_1^2 + \frac{\alpha^2}{4} \sqrt{ \frac{e_2}{e_1}  } \frac{1}{\sigma^2} \approx 0 \: , \\
\chi_{e_2 } &:=\:    P_2^2 + \frac{\alpha^2}{4} \sqrt{ \frac{e_1}{e_2}  } \frac{1}{\sigma^2} \approx 0 \: , \\
\chi_{\lambda_1 } &:=\:   X_1^2 + R^2  \approx 0 \: , 
\label{28c} \\
\chi_{\lambda_2 } &:=\:   X_2^2 + R^2  \approx 0  \: ,
\label{28d}
\end{align}
\end{subequations}
where $\chi$'s are constraint quantities for secondary constraints.
%%%
The $\tau$ evolution of $\chi$'s are obtained as 
\begin{subequations}
\begin{align}
%
% 1st line 
%
\dot{\chi}_{e_1} 
&=  -  \frac{\alpha^2}{8\sigma^2} \sqrt{ \frac{ e_2 }{ e_1 } } \left( \frac{ u_{e_1} }{e_1} - \frac{ u_{e_2} }{e_2} + \frac{4R}{\sigma K} \varepsilon_{-} \right) 
%- 4 \lambda_1 P_1 \cdot X_1  
\notag 
\\
&  
\quad  \,
- 4 \lambda_1 P_1 \cdot X_1  \: , 
\label{chi1} 
\\
%
% 2nd line
%
\dot{\chi}_{e_2}  
&=  \frac{\alpha^2}{8\sigma^2} \sqrt{ \frac{ e_1 }{e_2} } \left( \frac{ u_{e_1} }{e_1} - \frac{ u_{e_2} }{e_2} +  \frac{4R}{\sigma K} \varepsilon_{-} \right) 
\notag \\
\label{chi2} 
&\quad  \, 
- 4 \lambda_2 P_2 \cdot X_2  
\: , \\
\label{chi3} % 3rd line
\dot{\chi}_{\lambda_1} &= - 2 e_1 P_1 \cdot X_1  \: , \qquad
\\ 
\label{chi4} % 4th line
\dot{\chi}_{\lambda_2} &= - 2 e_2 P_2 \cdot X_2  \:   ,
\end{align}
\end{subequations}
where $ K :=  \sqrt{ R^4 - (X_1 \cdot X_2)^2 }  $ and $
\varepsilon_{-} := e_1 P_1 \cdot X_2 - e_2 P_2 \cdot X_1  .  $

From the consistency conditions for Eqs.~\eqref{chi3} and ~\eqref{chi4}, two more constraints are required:
%%%%%
\begin{align} \label{30}
\chi_5 :=  \: P_1 \cdot X_1 \approx 0  \:\: ,  \qquad \chi_6 :=  \: P_2 \cdot X_2 \approx 0  \:\: .
\end{align}
%%%%%
Together with these, the consistency conditions $ \dot{\chi}_{e_1} \approx 0 $ and $ \dot{\chi}_{e_2} \approx 0 $ become identical and give the single condition 
%%%%%
\begin{align}
u_{-} := \frac{u_{e_1}}{e_1} - \frac{u_{e_2}}{e_2} =   - \frac{ 4 R }{\sigma K} \varepsilon_{-}
\:\: .
\end{align}

The consistency conditions $\dot{\chi}_5 \approx 0 $ and $\dot{\chi}_6 \approx  0 $ give further secondary constraints
%%%%%
\begin{align}
\chi_7 :=& \:  e_1 P_1^2 + \frac{\alpha^2}{2} \sqrt{ e_1 e_2 } \frac{ R X_1 \! \cdot X_2 }{ \sigma^3 K } + 2 \lambda_1 X_1^2 \approx 0 \:\: , \\
\chi_8 :=& \:   e_2 P_2^2 + \frac{\alpha^2}{2} \sqrt{ e_1 e_2 } \frac{ R X_1 \! \cdot X_2 }{ \sigma^3 K } + 2 \lambda_2 X_2^2 \approx 0  \:\: .
\end{align}
From Eq.~\eqref{secondary} and these equations, it can be shown that 
\begin{align}
X_1 ^2 \, \approx X_2^2 \approx - R^2  \: , 
\qquad 	
\lambda_1 \, \approx \lambda_2  \: .
\end{align}
Using these weak equalities, we find that the consistency conditions $\dot{\chi}_7 \approx 0 $ and $ \dot{\chi}_8 \approx 0 $ determine $u_{\lambda_1}$ and $u_{\lambda_2}$ as 
\begin{align}
u_{\lambda_1} = u_{\lambda_2} = -  \frac{ \alpha^2  \sqrt{ e_1 e_2 }}{ 4 R \sigma^3 K} \Big( F+ 1 \Big) \varepsilon_{+}
		 + \frac{1}{4} \lambda_+ u_+ \: ,
\end{align}
%%%%%
where 
%%%%%
\begin{align}
\varepsilon_{+} := & \, \:   e_1 P_1 \cdot X_2 + e_2 P_2 \cdot X_1 
	\: , \\
F := & \, \:  1 + \frac{ ( X_1 \cdot X_2 )^2}{ K^2 } + \frac{ 3 R X_1 \cdot X_2 }{ \sigma K } 
	\: , \\
\lambda_{+} := & \,\: \lambda_1 + \lambda_2 \: , \\
u_{+} :=& \,\:  \frac{u_{e_1}}{e_1} + \frac{u_{e_2}}{e_2}  
	\: .
\end{align}
Thus, no new constraints are derived at this point.
While the multipliers $u_{-} ,\: u_{\lambda_1}$, and $u_{\lambda_2}$ are determined, $u_{+}$ remains undetermined due to the $\tau$ reparameterization invariance of the system.

To classify all the constraints into first- and second-class, we consider the following set of linear combinations of the constraint functions $\psi_a \; (a = 1,2 , \ldots ,12)$:
\begin{subequations}
\begin{align}
%  2nd - 1 
\label{psi_1}
\psi_1 & =  e_1 \phi_{e_1} - e_2 \phi_{e_2} + \frac{ \mathcal{P} }{ 2 R^2 \lambda_+ } ( \lambda_1 \phi_{\lambda_1} - \lambda_2 \phi_{\lambda_2} )
\: ,   \\
%%%%%%  2nd - 2 
\label{psi_2}
\psi_2 &= e_1 \chi_{e_1} - e_2 \chi_{e_2} - \frac{\mathcal{P} }{ R^2 \lambda_+ } ( \lambda_1 \chi_{\lambda_1} - \lambda_2 \chi_{\lambda_2} )
 \notag \\
 & \qquad 
 + \frac{ \alpha^2 \sqrt{ e_1 e_2 } \varepsilon_- F}{ R \lambda_+ \sigma^3 K } ( \lambda_1 \phi_{\lambda_1} + \lambda_2 \phi_{\lambda_2} )
\: , \\
%%%%%%  2nd - 3
\label{psi_3}
\psi_3 &= \lambda_1 \phi_{\lambda_1} + \lambda_2 \phi_{\lambda_2}
\: , \\
%%%%%%  2nd - 4 
\label{psi_4}
\psi_ 4 &= \chi_7 + \chi_8 + \frac{ \alpha^2 \sqrt{ e_1 e_2 } \varepsilon_- F}{ R \lambda_+ \sigma^3 K } ( \lambda_1 \phi_{\lambda_1 } - \lambda_2 \phi_{\lambda_2} ) \: ,
\\
%%%%%%  2nd - 5 
\label{psi_5}
\psi_5 &= \lambda_1 \phi_{\lambda_1 } - \lambda_2 \phi_{\lambda_2}
\: ,  \\
%%%%%%  2nd - 6 
\label{psi_6}
\psi_6 &= \chi_7 - \chi_8
\: ,  \\
%%%%%%  2nd - 7 
\label{psi_7}
\psi_7  &= \lambda_1 \chi_{\lambda_1}  +  \lambda_2  \chi_{\lambda_2}
\: ,   \\
%%%%%%  2nd - 8 
\label{psi_8}
\psi_8 &=  \left\{ \frac{ \alpha^2 \sqrt{e_1e_2} ( F+1 ) X_1 \cdot X_2}{R \lambda_+ \sigma^3 K} - 4 \right\} ( \lambda_1 \phi_{\lambda_1 } + \lambda_2 \phi_{\lambda_2} ) 
\notag \\
& \qquad 
+ \chi_5 + \chi_6 \: , 
\\
%%%%%%  2nd - 9
\label{psi_9}
\psi_9 &=  \lambda_1 \chi_{\lambda_1} - \lambda_2 \chi_{\lambda_2}
\: ,  \\
%%%%%%  2nd - 10
\label{psi_10}
\psi_{10} &= \chi_5 - \chi_6  - \left( \frac{ \mathcal{P} }{R^2 \lambda_+ } + 2 \right) ( \lambda_1 \phi_{\lambda_1 } - \lambda_2 \phi_{\lambda_2 } )
\: , 
%%%%%%  1st - 1 
\end{align}
\begin{align}
\label{psi_11}
\psi_{11} &=  e_1 \phi_{e_1} + e_2 \phi_{e_2} + \lambda_1 \phi_{\lambda_1} + \lambda_2 \phi_{\lambda_2}
\: , \\
%%%%%%  1st - 2 
\label{psi_12}
\psi_{12} &=
					e_1 \chi_{e_1} + e_2 \chi_{e_2} - 2 ( \lambda_1 \chi_{\lambda_1} + \lambda_2 \chi_{\lambda_2} ) 
					\notag \\
					& \quad 
						+ \frac{ 4R \varepsilon_- }{\sigma K} ( e_1 \phi_{e_1} - e_2 \phi_{e_2} ) 
					\notag \\
					& \quad
						+ \frac{ \alpha^2 \sqrt{ e_1 e_2 } \varepsilon_+ (1+F) }{ R \lambda_+ \sigma^3 K} ( \lambda_1 \phi_{\lambda_1} + \lambda_2 \phi_{\lambda_2} ) \: , 
\end{align}
%%%%%
with $ \mathcal{P} := e_1 P_1^2 + e_2 P_2^2 \: . $
%%%%%
\end{subequations}
The Poisson brackets between the new constraint functions form the $12\times 12$ matrix 
\begin{align}
\{ \psi_a , \psi_b \}
= 
\begin{pmatrix}
C & 0 \\
0 & 0
\end{pmatrix} 
\: , 
\end{align}
where $C = (C_{ab}) \: ( a,b= 1,2 \ldots , 10) $ is given by
\begin{align*} 
C= 
\begin{pmatrix} 
C_1 & & & & \\
 & C_2 & & & \\
 & & C_2 & & \\
 & & & C_2 & \\
 & & & & C_2
\end{pmatrix}
\end{align*}
%%%%%
with
%%%%%
\begin{align*}
C_1 = \begin{pmatrix} 0 & 2 V_{\mathrm{dS}} \\ - 2 V_{\mathrm{dS}} & 0  \end{pmatrix}
\: , 
\quad
C_2 =  \begin{pmatrix} 0 & - 2R^2 \lambda_+ \\ 2 R^2 \lambda_+ &  0  \end{pmatrix} 
\: ,
%}
\end{align*}
where $V_{\mathrm{dS}} $ has been defined in Eq.~\eqref{V}.
%%%%%
Therefore, the constraints $\psi_{11} \approx 0 $ and $\psi_{12} \approx 0 $ are classified as first class and the constraints $\psi_a \approx 0 \; ( a = 1,2, \ldots ,10 )$ are classified as second class.
%%%%%
The inverse matrix $C^{-1} = [( C^{-1} )^{ab} ] $ of $C$ reads as
%%%%%
\begin{align*}
C^{-1}= 
\begin{pmatrix} 
C_1^{-1} & & & & \\
 & C_2^{-1} & & & \\
 & & C_2^{-1} & & \\
 & & & C_2^{-1} & \\
 & & & & C_2^{-1}
\end{pmatrix} \: .
\end{align*}
%%%%%
Using this inverse matrix, we can define the Dirac bracket $ \{ F , \, G \}_{\mathrm{D}}$ between any functions of canonical variables as 
%%%%%
\begin{align} \label{defDB}
\Big\{ F \, , \, G \Big\}_{\mathrm{D}}	
&:= 
	\Big\{ F ,G \Big\}
	- \Big\{ F , \psi_a \Big\} (C^{-1})^{ab} \Big\{ \psi_b  , G \Big\}
\notag \\
&\: =
 	\Big\{ F \, , \, G \Big\} 
 		+ \frac{1}{2 V_{\mathrm{dS}}} \bigg(  
 			  \Big\{ F , \psi_1 \Big\} \Big\{ \psi_2 , G \Big\} 
 			- \Big\{ F , \psi_2 \Big\} \Big\{ \psi_1 , G \Big\} 
					  \bigg)
\notag \\
 & \qquad	- \frac{1}{2 R^2 \lambda_+ } \bigg(
			  \Big\{ F \, , \, \psi_3 \Big\} \Big\{ \psi_4 \, , \, G \Big\} - \Big\{ F \, , \, \psi_4 \Big\} \Big\{ \psi_3 \, , \, G \Big\} 
\notag \\
 &\quad \quad + \Big\{ F \, , \, \psi_5 \Big\} \Big\{ \psi_6 \, , \, G \Big\} - \Big\{ F \, , \, \psi_6 \Big\} \Big\{ \psi_5 \, , \, G \Big\} 
\notag \\
 &\quad \quad + \Big\{ F \, , \, \psi_7 \Big\} \Big\{ \psi_8 \, , \, G \Big\} - \Big\{ F \, , \, \psi_8 \Big\} \Big\{ \psi_7 \, , \, G \Big\}
\notag \\
 &\quad \quad + \Big\{ F \, , \, \psi_9 \Big\} \Big\{ \psi_{10} \, , \, G \Big\} - \Big\{ F \, , \, \psi_{10} \Big\} \Big\{ \psi_9 \, , \, G \Big\} 
			  								\bigg)
\: .
\end{align}

The nonvanishing Dirac brackets between the canonical variables are
%%%%%
\begin{subequations} \label{D}
\begin{align}
\Big\{  X_{ i }^{ A }  ,  \;  P_{ i }^{ B }  \Big\}_{\mathrm{D}} &= \eta^{ AB }  -  \frac{ 1 }{ X_i^2 } X_{ i }^{ A } X_{ i }^{ B } \: \: , \\
\Big\{  X_{ i }^{ A }  ,  \;  e_j  \Big\}_{\mathrm{D}} &= \frac{1}{V_{\mathrm{dS}}} \left(  \delta_{ i j } - \varepsilon_{ i j }  \right) P_{ i }^A e_{ i } e_{ j } \: \: , \\
%%%%%
\Big\{  P_{ i }^{ A }  ,  \;  e_j  \Big\}_{\mathrm{D}} &=  \frac{ \mathcal{P} }{ 2V_{\mathrm{dS}} R^2 } \left(  \delta_{ i j } - \varepsilon_{ i j }  \right) X_{ i }^A  e_{ j } \: \: , \\
\Big\{  e_{ i }  ,  \;  P_{ e_j }  \Big\}_{\mathrm{D}} &=  \delta_{ i j } +  \frac{ 1 }{ 2V_{\mathrm{dS}} } \left(  \delta_{ i j }  -  \sigma {}_{ i j }  \right)  e_{ i } P_{ j }^2  \: \: , \\
\Big\{  e_{ i }  ,  \;  \lambda_{ j }  \Big\}_{\mathrm{D}} &= (-1)^{i} \frac{ F \varepsilon_{ - } }{  KR \sigma  }  \delta_{ i j } e_{ i }  
%%%%%
\end{align}
\end{subequations}
with no summation over $i, j$, where $\sigma_{11} = \sigma_{22} = 0 , \: \sigma_{12} = \sigma_{21} = 1 $ and $ \varepsilon_{ij}= - \varepsilon_{ji} $ with $\varepsilon_{12} = 1$.

As long as the Dirac bracket is used, the second-class constraints can be set to zero strongly.
Thus, the first-class constraint $\psi_{12} \approx 0$ becomes
%%%%%
\begin{align}\label{cl_waveeq}
\psi_{12} 
	 &= e_1 \chi_{e_1} +  e_2 \chi_{e_2} \notag \\
	 &=  e_1 P_1^2 + e_2 P_2^2   + \frac{ \alpha^2 }{ 2 }  \frac{ \sqrt{e_1 e_2} }{ \sigma^2 }  \approx 0 \: .
\end{align}

The second-class constraints \eqref{psi_2} yields 
%%%%%
\begin{align}
\label{psi_2str}
e_1^2 P_1^4 + e_2^2 P_2^4 &= 2 e_1 e_2 P_1^2 P_2^2 \: .
\end{align}
%%%%%
Using this equation, the first-class constraint~\eqref{cl_waveeq} yields
%%%%%
\begin{align} \label{lambda}
\varLambda := P_1^2 P_2^2 - \frac{\alpha^4}{16} \frac{ 1 }{ \sigma^4 } &\approx 0 \: .
\end{align}
%%%%%%
%%%This constraint is precisely a de Sitter version of one derived in Ref. \cite{Casalbuoni2014c}.

\section{New canonical variables}

As seen from Eq.~\eqref{D}, the pairs $(X^{A}_{i}, P_{i A})$, $(e_{i}, P_{e_{i}})$ and $(\lambda_{i}, P_{\lambda_{i}})$, which consist of the 28 canonical variables, are no longer canonically conjugate pairs under the Dirac bracket.
This is due to the 10 second-class constraints and we have only 18 independent degrees of freedom that may constitute canonical pairs in the sense of Dirac brackets.
To identify them, we define new variables as follows:
%%%%%
\begin{subequations} \label{newvariables}
\begin{align}
 {x}^{\mu}_{i} &:= X^{\mu}_{i}, & 
 {p}_{i \mu}  &:= g_{\mu \nu}(x_{i}) P^{\nu}_{i}, & \label{tilde2p}
\\
 {e} &:= \frac{ e_1 e_2 }{2} ,  &
 {\pi} &:= \frac{ e_1 P_{e_1} + e_2 P_{e_2} }{ e_1 e_2} , &\label{tilde4pi}
\end{align}
\end{subequations}
%%%%%%
with 
\begin{align} \label{metric}
 g_{\mu \nu}(x_{i}) := \eta_{\mu \nu} -\frac{ x_{i \mu} x_{i \nu} }{ \eta_{\alpha \beta} x^{\alpha}_{i}  x^{\beta}_{i} +R^{2} }  \; ,
\end{align}
where $\mu, \nu = 0, 1,2,3$ and $ x_{i\mu} := \eta_{\mu \nu} x^{\nu}_i $.
%%%%%%
By using Eq.~\eqref{D}, it is easily checked that the nonvanishing Dirac brackets between the new variables become
\begin{align}\label{newDB}
 \left\{ {x}^{\mu}_{i}, {p}_{j \nu} \right\}_{\mathrm{D}} &=
\delta_{ij} \delta^{\mu}_{\nu} \; , &
 \left\{ e , \; \pi  \right\}_{\mathrm{D}} &= 1 \: \: . &
\end{align}
Therefore each pair of $({x}^{\mu}_{i}, {p}_{i \mu}) $ and $  (e , \pi)$ forms a pair of canonical variables in the sense of Dirac brackets.

It is noted that taking the pairs $(x_i^{\mu}, p_{i \mu} )$ as canonical pairs corresponds to choosing a particular coordinate system $(x^{\mu})$ for the de Sitter space.
To see this, we solve Eqs.~\eqref{28c} and \eqref{28d} with respect to $X^{5}_{1}$ and $X^5_{2}$, respectively, and write them in terms of $x^{\mu}_{i}$ as
\begin{align} 
 X_i^5(x_{i}) &= \pm \sqrt{ \eta_{\mu \nu} X^{\mu}_i  X^{\nu}_i + R^2 } = \pm \sqrt{ \eta_{\mu \nu} x^{\mu}_i  x^{\nu}_i + R^2 } \: . \label{X5} 
\end{align}
%%%%%%
This is the same as the parameterization of the de Sitter space \eqref{local} and Eq.~\eqref{metric} is the induced metric \eqref{induced}.
By solving Eq.~\eqref{30} with respect to $P^5_i$, we get  
\begin{align}
 P_i^5 ({x}_{i}, p_{i}) &= \frac{ \eta_{\mu \nu} X^{\mu}_i  P^{\nu}_i }{ X_i^5} = \pm \frac{ g^{\mu \nu} (x_{i}) x_{i \mu} p_{i \nu} }{ \sqrt{ \eta_{\mu \nu} x^{\mu}_i  x^{\nu}_i + R^2 } } \: ,
\end{align}
where 
$  g^{\mu \nu}(x_{i}) = \eta^{\mu \nu} +  x^{\mu}_{i} x^{\nu}_{i} / R^{2}    $ is the inverse of $g_{\mu\nu} (x) $ at $x= x_i$.

Using new canonical variables \eqref{newvariables}, the first-class constraint $\psi_{11} \approx 0 $ is written as $ e \pi \approx 0$, which is   equivalent to $ \pi \approx 0  $.
Together with Eq.~\eqref{lambda}, the first-class constraints are now listed as
\begin{subequations} \label{1st}
\begin{align} \label{1st1}
\pi &\approx 0 \: , 
\\
\label{1st2}
 \varLambda (x_{1},x_{2},p_{1},p_{2}) = P_1^2 P_2^2 - \frac{\alpha^4}{16} \frac{ 1 }{ \sigma^4(x_{1},x_{2}) } &\approx 0 \:  \: ,
\end{align}
\end{subequations}
where $P_i^2$ and $\sigma (x_{1}, x_{2})$ are written as 
\begin{align} \label{P^2}
 P_i^2 &= 
 g^{\mu \nu}(x_{i}) p_{i\mu} p_{i\nu}  \:, \\
\sigma (x_{1}, x_{2}) &=  R \mathrm{cos}^{-1} \left( - \frac{  \eta_{\mu \nu} x^{\mu}_1  x^{\nu}_2 +  X^5_1(x_1) X^5_2(x_2)  }{R^2} \right) \: .
\end{align}

In the next sectoin, we carry out the canonical quantization of our system based on   the Dirac brackets \eqref{newDB} for the new variables and the first-class constraints \eqref{1st}.

\section{Canonical quantization}

We set the commutation relations between $\hat{f}$ and $\hat{g}$ corresponding to functions $f$ and $g$ of canonical variables, in such manner
\begin{align}
 [ \hat{f}   \, , \: \hat{g} ] =   i \widehat{ \{ f  , \: g \}}_{\mathrm{D}}
\end{align}
in units $\hbar=1$. The symbol $\widehat{ \{ f  , \: g \}}_{\mathrm{D}}$ denotes the operator corresponding to the Dirac brackets $\{ f , g \}_{\mathrm{D}}$.
From Eq.~\eqref{newDB},  we have the commutation relations 
\begin{align}  \label{ccr}
\big[ \hat{x}_i^{\mu}, \;  \hat{p}_{j \nu} \big] &=  i \delta_{ij} \delta^{\mu}_{\nu} \: , &
\big[ \hat{e}  , \; \hat{ \pi }  \big]  &=  i \: . &
\end{align}

We consider the quantum mechanics characterized by the above commutation relations in the coordinate representation, where the configuration space is the de Sitter space endowed with the metric \eqref{induced}.
The coordinate representation for curved configuration spaces was considered by DeWitt \cite{Dewitt1952}.
%%%%%%
Following DeWitt, we take the simultaneous eigenstates of $\hat{x}_i , \; \hat{e}$
\begin{align} \label{eigen}
\hat{x}^{\mu}_i  | x_1 , x_2  , e \rangle &=  x_i^{\mu}   | x_1 , x_2  , e \rangle  \: , 
\\
\hat{e}   | x_1 , x_2  , e \rangle &= e  | x_1 , x_2  , e \rangle \: 
\end{align}
as a basis of the Hilbert space.
%%%%%%
Then we expand an arbitrary state $ | \varphi \rangle $ in terms of the basis vectors in the form 
\begin{align}
| \varphi \rangle  =  \int \Big( \prod_{i=1,2} \sqrt{ - g(x_i) } d^4 x_i \Big) de \,   \varphi(x_1,x_2 ,e )  | x_1,x_2 , e  \,  \rangle    \:  ,
\end{align}
with $ g(x) = \det g_{\mu \nu} (x) $ and wave functions $  \varphi(x_1,x_2 ,e )  \equiv \langle x_1, x_2 ,e  |  \varphi \rangle $.
In addition, we assume that the eigenvectors $| x_1,x_2 , e  \,  \rangle$ satisfy the normalization condition 
\begin{align} \label{normalization}
\langle x''_1 , x''_2 , e''  | x'_1 , x'_2 , e' \rangle  
=  \delta (e'' - e' )  \prod_{i=1,2} \frac{  \delta^{4} (x_i'' - x_i ')  }{\sqrt{ - g(x_i') } }  \: ,
\end{align}
which is consistent with the completeness relation  
\begin{align}
1 =  \int \prod_{i=1,2} \sqrt{ - g(x_i) } d^4 x_i de \,  | x_1,x_2,e \rangle \langle x_1, x_2, e |  \:\: .
\end{align}
%%%%%%
On the basis vectors satisfying Eq.~\eqref{normalization}, the momentum operators $\hat{p}_i{}_{\mu}$ and $ \hat{\pi}$ can be represented \cite{Dewitt1952} as 
\begin{align} \label{p_rep}
\langle x_1 , x_2 , e  |  \hat{p}_{i \mu} &= \left( - i \frac{\partial }{\partial  x^{\mu}_i } - \frac{i}{2} \Gamma_{\mu} (x_i)  \right) \langle x_1 , x_2 , e  |  \: , 
\\
\langle x_1 , x_2 , e  | \hat{\pi} &= - i \frac{\partial }{\partial e} \langle x_1 , x_2 , e  | \: ,
\label{reppi}
\end{align}
%%%%%%
where $ \Gamma_{\mu} (x) :=  \Gamma^{\nu}_{\nu \mu } $ and the Christoffel symbol $ \Gamma^{\sigma}_{\mu \nu } $ is defined as
\[
 \Gamma^{\sigma}_{\mu \nu } 
 =  \frac{1}{2} g^{\sigma \tau}  
 \left( 
  \frac{\partial}{\partial x^{ \nu }}  g_{ \tau \mu  }   
  +  \frac{\partial}{\partial x^{ \mu }}  g_{\tau \nu}   
 - \frac{\partial}{\partial x^{ \tau }}  g_{\mu \nu}   
 \right) \: .
\]
For confirming the hermiticity of $ \hat{p}_{i \mu } $, namely $\langle \psi | \hat{p}_{i \mu }  \varphi \rangle = \langle \hat{p}_{i\mu} \psi | \varphi \rangle $, it is useful to use an alternative expression 
%%%%%%
\begin{align} \label{hermite}
\langle x_1 , x_2 , e  |  \hat{p}_{i \mu} 
= (-g(x_i))^{-\frac{1}{4} } \left( -i \frac{\partial }{\partial x_i^{\mu} } \right) (-g(x_i))^{\frac{1}{4}}   \langle x_1 , x_2 , e  |  \: .
\end{align}
%%%%%%
This is also convenient to verity that the Laplace-Beltrami operator $\square^{(dS)}_i$ at a point $x_i^{\mu}$ on ${\rm dS}_4 $ can be written as
%%%%%%
\begin{align}
\square^{(dS)}_i  &:=  \frac{1}{ \sqrt{ - g (x_i) } } \frac{\partial }{\partial x_i^{\mu}} \sqrt{ -g (x_i) } g^{\mu \nu } (x_i) \frac{\partial }{\partial x_i^{\nu} }
\notag 
\\
&= -
(-g(x_i))^{- \frac{1}{4} }  \hat{p}_{i \mu} \sqrt{-g(x_i)} g^{\mu \nu }(x_i ) \hat{p}_{ i \nu } ( - g(x_i) )^{ - \frac{1}{4} } \:  \: 
\label{box}
\end{align}
%%%%%%
in the $x$-representation.

Now, let us derive a wave equation associated with the first class constraints.
%%%%%%
We replace Eqs.~\eqref{1st1} and \eqref{1st2} with the conditions that single out the physical states, respectively:
%%%%%%
\begin{subequations} \label{PhysCond}
\begin{align}  \label{PhysCond0}
\hat{\pi} \, | \varphi \rangle & = 0  \:  ,  \\
\label{PhysCond1}
\hat{\varLambda} ( \hat{x}_1, \hat{x}_2 , \hat{p}_1 , \hat{p}_2  ) \, 
 | \varphi \rangle & = 0   \: \: , 
\end{align}
\end{subequations}
%%%%%%
where
%%%%%%
\begin{align} \label{defLambda}
\hat{\varLambda} ( \hat{x}_1, \hat{x}_2 , \hat{p}_1 , \hat{p}_2  )   := \widehat{P_1^2 } \widehat{ P_2^2 } - \frac{ \alpha^4 }{16 \sigma^4 (\hat{x}_1,\hat{x}_2 )  }  \: \: ,
\end{align}
%%%%%%
and $ \widehat{ P_i^2  }  $ is the quantum mechanical operator corresponding to the classical quantity $ P_i^2 = g^{\mu \nu }(x_i) p_{i \mu } p_{i \nu} $ in Eq.~\eqref{P^2}. 
Obviously, there is ordering ambiguity in determining an explicit expression of $ \widehat{ P_i^2  } $.
%%%%%%%%%%
Since $ P_i^2 $ is a scalar under general coordinate transformations, we simply require $ \widehat{P_i^2} $ to be also a scalar for a particular ordering of $ \widehat{P_i^2} $.
\footnote{The Weyl ordering rule, commonly taken in canonical quantization procedure, does not satisfy our requirement because the Weyl ordered operator, $ \widehat{P_i^2}_{(W)} $, for $P_i^2 = g_{\mu \nu} p^{\mu}_i p^{\nu}_i $ is explicitly evaluated as 
\begin{align*}
 \widehat{P_i^2}_{(W)}  &=  \hat{p}_i^{\mu} g_{\mu \nu} (x_i ) \hat{p}^{\nu}_i - \frac{1}{4} \frac{ \partial g_{\mu \nu}(x_i) }{ \partial x_i^{\mu} \partial x_i^{\nu} }
 \\
 &= - \square_i - \frac{1}{4} \mathcal{R}(x_i) - \frac{1}{4} g^{\mu\nu}  (x_i) \Gamma^{\alpha}_{\beta \mu} \Gamma^{\beta}_{\alpha \nu }(x_i) 
 \: 
\end{align*}
with no sum over $i$, where $\mathcal{R}(x)$ is the Ricci scalar.
The last term in the last line is not invariant under general coordinate transformations.
}
%%%%%%%%%%
\begin{align}  \label{ordering}
\widehat{ P_i^2  } 
:=
(-g(\hat{x}_i))^{- \frac{1}{4} }  \hat{p}_{i \mu} \sqrt{-g(\hat{x}_i)} g^{\mu \nu }(\hat{x}_i ) \hat{p}_{ i \nu } ( - g(\hat{x}_i) )^{ - \frac{1}{4} } \: .
\end{align}
%%%%%%
On a bra vector, this is represented as
%%%%%%
\begin{align} \label{box2}
\langle x_1 , x_2 , e  | \widehat{ P_i^2  }  = - \square^{(dS)}_i \langle x_1 , x_2 , e  | \:\: .
\end{align}
%%%%%%
Using Eq.~\eqref{defLambda}  and Eq.~\eqref{box2}, we have 
\begin{align}  \label{repLambda}
&\langle x_1 , x_2 , e  | \hat{\varLambda} ( \hat{x}_1, \hat{x}_2 , \hat{p}_1 , \hat{p}_2  )   \notag 
\\
&= 
\left( \square^{(dS)}_{1} \square^{(dS)}_{2}  - \frac{\alpha^4 }{16 \sigma^4(x_1,x_2)} \right) \langle x_1 , x_2 , e  | \: .
\end{align}
%%%%%%
Multiplying Eqs.~\eqref{PhysCond0} and \eqref{PhysCond1} by $\langle x_1, x_2, e |$ and using Eqs.~\eqref{reppi} and \eqref{repLambda}, respectively,  we have simultaneous partial differential equations for wave functions $\varphi(x_1,x_2,e)$:
%%%%%%
\begin{subequations}
\begin{align}
\frac{\partial}{\partial e} \varphi( x_1, x_2, e ) &= 0 \: ,
\label{simuleq1}
\\
\left( \square^{(dS)}_{1} \square^{(dS)}_{2}  - \frac{\alpha^4 }{16 \sigma^4(x_1,x_2) } \right)
\varphi (x_1,x_2 ,e) &= 0  \: .
\label{simuleq2}
\end{align}
\end{subequations}
%%%%%%
Eq.~\eqref{simuleq1} tells us that the wave functions do not depend on the variable $ e$.
Therefore we can write $\varphi(x_1,x_2 ,e) = \varphi (x_1 , x_2)$ and  Eq.~\eqref{simuleq2} becomes the wave equation for bilocal fields $\varphi(x_1,x_2)$:
%%%%%%
\begin{align} \label{waveeq}
\left( \square^{(dS)}_{1} \square^{(dS)}_{2}  - \frac{\alpha^4 }{16 \sigma^4(x_1,x_2) } \right)
\varphi (x_1,x_2) = 0  \: .
\end{align}
%%%%%%
This is precisely a de Sitter version of the wave equation \eqref{waveeq-cg} originally derived for the case of Minkowski space in Ref. \cite{Casalbuoni2014c}.
The squared geodesic distance has the property $\sigma^2 \rightarrow -r^2$ as  $R \rightarrow \infty$ as shown in Sec.~2, so that Eq.~\eqref{waveeq}  also reduces to Eq.~\eqref{waveeq-cg}.

Although it is important to consider the well-posedness of an initial value problem for the wave equation \eqref{waveeq}, further investigations for it do not enter into the scope of this letter.

\section{Conclusion}

In this letter, we have generalized the conformally invariant action $S$, Eq.~\eqref{S}, for two interacting massless particles in Minkowski space to the case of de Sitter space from the perspective of geodesic distance.
To begin with, we have defined the action $S_g$, Eq.~\eqref{S'}, as a generalization of $S$ to general spacetimes that become close to Minkowski space in a certain limit.
%%%%%%
After that, we have specified the action $S_{\mathrm{dS}}$, Eq.~\eqref{SdS},  as $S_g$ for the case of de Sitter space in terms of the embedding coordinates $X_i^A$ with the explicit expression of the geodesic distance \eqref{sigma} and also confirmed that $S_{\mathrm{dS}} $ really reproduces $S$ in the limit  $R \rightarrow \infty$ with the coordinate chart \eqref{local}.
%%%%%%
In the Hamilton analysis, according to Dirac's prescription \cite{Dirac:1964:LQM} for the constrained Hamiltonian systems, we have carried out the classification of the constraints and have defined the Dirac bracket from the second-class constraints.
%%%%%%
Then we have found that the new variables \eqref{newvariables} constitute the canonically conjugate pairs in the sense of Dirac brackets.
%%%%%%
Following DeWitt~\cite{Dewitt1952}, we have performed the subsequent canonical quantization of the system in coordinate representation where the configuration space is $\mathrm{dS}_4$ endowed with the metric \eqref{induced}.
By choosing the operator ordering \eqref{ordering} for $\widehat{P}_i^2 $, the physical state conditions \eqref{PhysCond} associated with the first-class constraints \eqref{1st} has been read as the fourth-order differential equation \eqref{waveeq} for bilocal fields $\varphi(x_1,x_2)$, which can be recognized as a de Sitter version of the wave equation \eqref{waveeq-cg} derived based on the action \eqref{S} in Ref.~\cite{Casalbuoni2014c}.

It seems worthwhile to quantize the model governed by the action \eqref{S'} in path integral formulation in curved spaces, which was studied for a nonrelativistic particle in Ref.~\cite{Cheng1972} and for a relativistic particle in a manifestly covariant way in Ref.~\cite{Bastianelli1991a}.

It is also interesting to consider the dynamics of bilocal fields satisfying the wave equation \eqref{waveeq}, including to confirm the well-posedness of \eqref{waveeq}, which may possibly be relevant to higher derivative theories of scalar fields in curved spacetime \cite{Gibbons2019a} and/or higher spin theories \cite{Sorokin2005}.

\section*{Acknowledgements}

The authors thank the members of the theoretical group in Nihon University for providing the research environment.
N.K. is also grateful to Kengo Maeda and Shigefumi Naka for their useful comments in the early stage of this work and appreciate kind support from Haruki Toyoda.
S.O. would like to thank Shinichi Deguchi for his insightful comments.

\bibliography{library,supplement}

\end{document}